\documentclass[prd,aps,onecolumn,preprintnumbers,amsmath,amssymb]{revtex4}
\usepackage{amsmath}
\usepackage{graphicx}
\usepackage{dcolumn}
\usepackage{bm}
\usepackage{amssymb}
\usepackage{latexsym}

\def\be{\begin{equation}}
\def\ee{\end{equation}}
\def\bea{\begin{eqnarray}}
\def\eea{\end{eqnarray}}

\begin{document}

\markright{CYCU-HEP-10-09}

\title{Can the Big Bang Singularity be avoided by a single scalar field?}

\author{Taotao Qiu\footnote{qiutt@mail.ihep.ac.cn}}

\affiliation{Physics Department, Chung-Yuan Christian University,
Chung-li, Taiwan 320}

\begin{abstract}
In this note, we investigate the possibility of avoiding the Big
Bang singularity with a single scalar field which couples
non-minimally to gravity. We show that in the case that gravity
couples linearly to the field, some severe conditions on the field's
potential have to be imposed. However, in non-linear case, it is
quite generic to avoid the singularity with single scalar field.
\end{abstract}

\maketitle

\section{introduction}

The Big Bang Singularity (BBS) is a notorious problem that present
in Standard Big Bang Theory as well as inflation theory
\cite{Hawking:1973uf}. People have made a lot of efforts in solving
this problem, and lots of mechanisms have been carried out, among
which there are scenarios from string theory, quantum gravity theory
and so on. Either reduced from these theories, or directly arisen
from the effective field theory, we are able to obtain a bounce
solution in 4-dimensional space-time. Bouncing cosmologies,
especially non-singular ones, has gained more and more development
in many aspects \cite{Novello:2008ra}. As an alternative of the
standard theory in the early universe, it can not only avoid BBS,
but also gives well favored observational effects compared to the
astronomical data, such as scale invariant power spectrum and
suppression of CMB low quardrupole on large scales
\cite{Piao:2003zm}. Besides, a scenario of bouncing cosmology can
give rise to large nongaussianities \cite{Cai:2008ed}.

If we take our universe to be flat, thus within the framework of
Einstein's General Relativity in 4 dimensional space-time, a bounce
is usually realized by Quintom matter \cite{Cai:2007qw}\footnote{Of
course there are many other alternatives which lead to a bounce,
such as non-scalar fields or non-perfect fluids
\cite{Brechet:2008zz}. We thank the anonymous referee to remind us
this.}. As has been briefly proved in \cite{Cai:2007qw}, in order
for a bounce to happen, where the scale factor of the universe
contracted to some value and then expanded, the Hubble parameter has
to evolve from a negative value to a positive one. It will cause the
equation of state of the universe will be less than $-1$, violating
the Null Energy Condition (NEC). Generally, some Phantom part will
get involved that makes the physics unclear \cite{Carroll:2003st}.

However, this dilemma can be ameliorated in many ways, such as to
introduce non-scalar fields, or higher derivative operators and so
on. One of the simplest ways is to go with the help of non-minimal
coupling to Gravity \cite{Abreu:1994fd}. In quantum gravity in
curved space-time, it is argued that the existence of the
non-minimal coupling term is required by the quantum corrections and
renormalization \cite{Chernikov:1968zm}. This coupling includes
scalar-tensor theories where the Ricci scalar couples to the field
through the term $F(\phi)R$ \cite{Fujii} (as will be called ``linear
coupling'' for convenience), which contains Brans-Dicke theory
\cite{Brans:1961sx} or dilaton theory
\cite{Gasperini:1992em,Bamba:2006mh}, see \cite{Sotiriou:2008rp} for
a comprehensive review. Since the inclusion of non-minimal coupling
term can drive cosmic acceleration with a wider class of potential
than usually considered, it can also be utilized as inflaton in the
early time \cite{Abbott:1981rg} and dark energy at current epoch
\cite{Uzan:1999ch}. With the help of the non-minimal coupling, a
single scalar field can also behave like systems of
multi-degree-of-freedom without really involving a ghost field, such
as having its equation of state (EoS) cross the cosmological
constant boundary as a ``Quintom" matter
\cite{Cai:2005ie}\footnote{For Quintom investigations please see
\cite{Feng:2004ad} and the following literature, one can also see
\cite{Cai:2009zp} for a review.}. Moreover, a great deal of
non-minimal coupling theories can find their equivalence to the
modified gravity theories \cite{Flanagan:2003rb}.

In this paper we study the possibility of realizing bouncing
scenario with a single scalar field non-minimally coupled to
Gravity. We find that generally speaking, it does work as expected,
except that for some specific case where Gravity couples linearly,
some constraints have to be imposed on the field action. The paper
is organized as follows: in section II we shortly demonstrate the
general case for a non-minimal coupling field to drive a bounce, in
section III and IV we investigate the linear coupling case and
non-linear coupling case respectively, while conclusions come in the
last section.

\section{general conditions for a single field to avoid Big Bang Singularity}
To begin with, let us consider the most general action containing
one scalar field and gravity, of which the two components are
coupled in a very general form: \be S=\int
d^{4}x\sqrt{-g}\{\frac{f(R,\phi)}{2\kappa^{2}}+P(X,\phi)\}~,\ee
where $\kappa^2\equiv 8\pi G$ and the metric used here is
$g_{\mu\nu}=diag[1,-a^2(t),-a^2(t),-a^2(t)]$. $X$ denotes the
kinetic term of the field:
$X\equiv\frac{1}{2}\nabla_\mu\phi\nabla^\mu\phi$. What we're working
throughout this note is under Jordan frame\footnote{As is
well-known, the non-minimal coupling can be got rid of by performing
the conformal transformation and redefining the scalar field. In
that case, a bounce may not occur. This is due to the physical
difference between Jordan and Einstein frames, in which the
evolutions of the universe are not the same. We want to clarify this
in some future work.}. By variation of the action we can easily
obtain the Friedmann Equations: \bea
\label{fe1} -\frac{f(R,\phi)}{2}-\kappa^{2}P(X,\phi)+\kappa^{2}P_{X}\dot{\phi^{2}}+3(\dot{H}+H^{2})f_{R}-3H\dot{f_{R}}&=&0~,\\
\label{fe2}
\frac{f(R,\phi)}{2}+\kappa^{2}P(X,\phi)-f_{R}(\dot{H}+3H^{2})+\ddot{f_{R}}+2H\dot{f_{R}}&=&0~,\eea
where $f_R\equiv\frac{df}{dR}$, and the equation of motion for the
field $\phi$ is: \be\label{eom}
-P_{X}\Box\phi-P_{XX}\nabla_{\mu}X\nabla^{\mu}\phi-P_{X\phi}\nabla_{\mu}\phi\nabla^{\mu}\phi+P_{\phi}+\frac{f_{\phi}}{2\kappa^{2}}=0~.\ee
where $\Box\phi=\ddot\phi+3H\dot\phi$ and similarly
$f_\phi\equiv\frac{df}{d\phi}$ is the derivative of $f(R,\phi)$ with
respect to the field $\phi$.

As mentioned before, for a flat universe with standard Einstein
Gravity in 4-dimensional space-time, a bounce happens only when the
Null Energy Condition (NEC) is violated. We can learn from Einstein
Equations that this leads to the vanishing of Hubble parameter $H$
with a positive time derivative, namely $H=0$ and $\dot{H}>0$ at the
bounce point \cite{Cai:2007qw}. Substituting the first condition to
the above equations one gets: \bea
&&3\ddot f_R+\kappa^2(\rho+3P)+f=0~,\\
&&\dot H=\frac{f+2\kappa^2\rho}{6f_R}>0~.\eea

In following sections, we will analyze in detail the application of
these conditions to the non-minimal coupling scalar field. Before
starting, we would like to classify the coupling term into two
categories according to its properties: one is a more specific case
that the Gravity couples linearly to the field, and the other is a
more general case which refers to non-linear coupling. Actually,
these two categories are very different due to the different numbers
of degree of freedom in the background equations. Due to this
reason, it may cause very different conclusions.

\section{Category I: Gravity coupling linearly to the field}
\subsection{general solution}
Among the various possible forms the coupling terms may have, it is
the simplest to study the one that Ricci scalar couples linearly to
the field, which is easy to get a renormalizable theory without any
suppression. Moreover, as differ from its non-linear counterpart, in
this case the higher order derivative of $R$ is zero, thus higher
order time derivatives of $H$ will not get involved into the
background equations. To study the general solution of this case,
let us assume that $f(R,\phi)=R(1-F(\phi))$, where the first term in
the bracket denotes the standard term from General Relativity, and
the second comes from the non-minimal coupling. From Eqs.
(\ref{fe1}) and (\ref{fe2}), one can get: \bea
-\frac{R(1-F(\phi))}{2}+\kappa^{2}\rho+3\dot{H}(1-F(\phi))&=&0~,\\
\frac{R(1-F(\phi))}{2}+\kappa^{2}P(X,\phi)-\dot{H}(1-F(\phi))\ddot{-F(\phi)}&=&0~,\eea
which can be furtherly simplified to a very neat form:
\bea -3H^{2}(1-F)+\kappa^{2}\rho+3H\dot{F}&=&0~,\\
2\kappa^{2}XP_{X}+2\dot{H}(1-F(\phi))-2XF_{\phi\phi}-F_{\phi}\ddot{\phi}&=&0~.\eea
In deriving these equations, we have made use of the fact
$\rho=2XP_X-P$ and $\dot X=\dot\phi\ddot\phi$. The last one of the
above equations gives \be
\ddot{\phi}=\frac{2\kappa^{2}XP_{X}+2\dot{H}(1-F(\phi))-2XF_{\phi\phi}}{F_{\phi}}~.\ee
For $H=0$, we have \bea \rho&=&0~,\\
\dot{H}&=&-\kappa^{2}\frac{\rho_{X}(2\kappa^{2}XP_{X}-2XF_{\phi\phi})+\rho_{\phi}F_{\phi}}{3F_{\phi}^{2}+2\kappa^{2}\rho_{X}(1-F)}~.\eea
In deriving the last equation we also made use of Eq. (\ref{eom}).
We can see from above that for case of gravity coupling linearly to
scalar field, a bounce requires the rhs of the above equation to be
larger than 0 as well as the energy density equals to 0. This can be
viewed as a general condition for this case.

\subsection{single scalar field with canonical form} At first let us
consider the simplest case in which the field has a standard
canonical form: \be\label{canonical}
P(X,\phi)=\frac{1}{2}\partial_\mu\phi\partial^\mu\phi-V(\phi)~,\ee
where $V(\phi)$ is its potential. Thus the energy density of the
field $\rho=\frac{1}{2}\dot\phi^2+V(\phi)$. The condition of
$\rho=0$ requires that at the bounce point
$\frac{1}{2}\dot\phi^2+V(\phi)=0$, or equivalently
$V(\phi)=-\frac{1}{2}\dot\phi^2$. It can be straightforwardly seen
that as long as $\dot\phi\neq 0$ at this point, one needs
$V(\phi)<0$, namely, a potential containing negative values at least
at some region is required, while a positive-definite potential as
usual doesn't work. But is that the case even for the possible case
where $\dot\phi=V(\phi)=0$, though it refers to a very specific
point in $(\phi,\dot\phi)$ space and needs strong fine-tuning? To
see this, let's take a look at the other condition of $\dot H>0$,
which then indicates: \be
\frac{V_{\phi}F_{\phi}}{3F_{\phi}^{2}+2\kappa^{2}(1-F)}<0~.\ee where
we used $\rho_\phi=V_\phi$ as well as $\rho_X=1$. From this equation
we notice that $V_\phi$ can be either larger or less than 0,
depending on the form of $F(\phi)$. For example, if we assume that
$F(\phi)=\kappa^2\xi\phi^2$, then the condition is $\xi\phi
V_\phi<0$. But anyway, as long as $V_\phi\neq0$ at the bouncing
point where $V(\phi)=0$, there will inevitably be some region where
$V(\phi)<0$. So this will be a generic constraint on the potential
of the field. In usual case, this will break down the
positive-definition of the energy density, which causes unphysics.
However, here it is fine both because of the positive kinetic term
and the effects of gravity. Thus, after some proper choice of
parameters, we can still get a self-consistent system.

As a further study, we also derived the second derivative of $H$
with respect to $t$ at the bouncing point. It appears to be: \bea
\ddot
H&=&\frac{1}{[2\kappa^{2}(1-F)+3F_{\phi}^{2}]^{2}}\{[2\kappa^{2}(1-F)+3F_{\phi}^{2}]
[\kappa^{2}(-2\kappa^{2}\dot{\phi}\ddot{\phi}-4\dot{H}\dot{\phi}F_{\phi}-4H\ddot{\phi}F_{\phi}-4H\dot{\phi}^{2}F_{\phi\phi}\nonumber\\
&&-V_{\phi\phi}F_{\phi}\dot{\phi}-V_{\phi}F_{\phi\phi}\dot{\phi}+F_{\phi\phi\phi}\dot{\phi}^{3}+2F_{\phi\phi}\dot{\phi}\ddot{\phi})
-12H\dot{H}F_{\phi}^{2}-12H^{2}F_{\phi}F_{\phi\phi}\dot{\phi}]\nonumber\\
&&+(2\kappa^{2}F_{\phi}\dot{\phi}-6F_{\phi}F_{\phi\phi}\dot{\phi})[\kappa^{2}(-\kappa^{2}\dot{\phi}^{2}-4H\dot{\phi}F_{\phi}
-V_{\phi}F_{\phi}+F_{\phi\phi}\dot{\phi}^{2})-6H^{2}F_{\phi}^{2}]\}~.\eea
It can be straightforwardly read off that for the
$\dot\phi=V(\phi)=0$ case (note that $H=0$), $\ddot H$ will vanish
at the bouncing point. This means that in this case, the velocity of
Hubble parameter always reaches its extreme value when passing
through the bouncing pivot. This is a new property that hasn't been
pointed out by other authors in the literature.

\begin{figure}[htbp]
\includegraphics[scale=0.3]{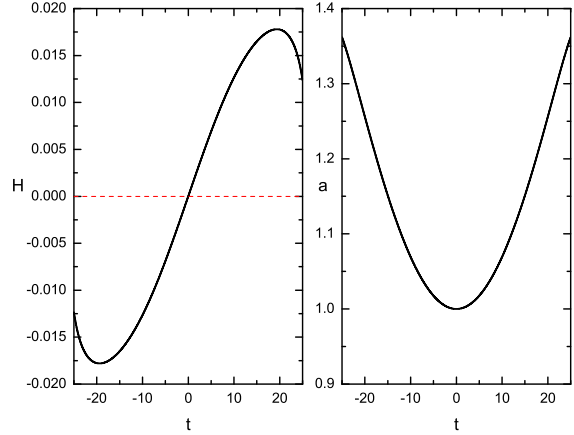}
\caption{The case of canonical single scalar field and linear
non-minimal coupling. The potential is chosen to be
$V(\phi)=\frac{1}{2}m^2\phi^2-V_0$ and the coupling term is
$RF(\phi)=\kappa^2\xi R\phi^2$. The parameter is chosen to be
$m=0.1m_{pl}$, $V_0=0.005m_{pl}^4$ and $\xi=-1.0$, and the initial
conditions are $\phi_i=0.473m_{pl}$,
$\dot\phi_i=0.061m_{pl}^2$.}\label{linear1}
\end{figure}

\begin{figure}[htbp]
\includegraphics[scale=0.3]{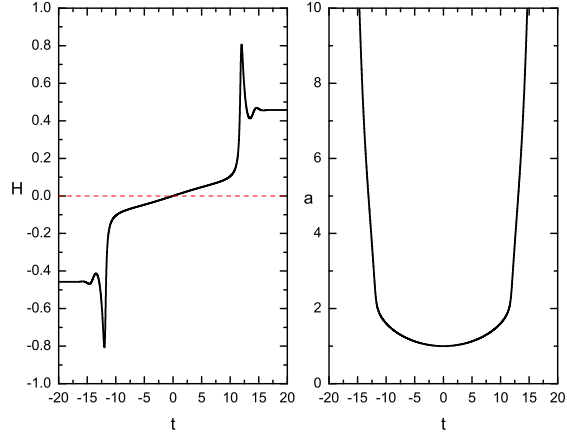}
\caption{The potential is chosen to be
$V(\phi)=V_0(e^{\lambda\phi^2}-\frac{1}{2})$ and the coupling term
is $RF(\phi)=\kappa^2\xi R\phi^2$. The parameter is chosen to be
$\lambda=-1.0m_{pl}^{-2}$, $V_0=0.05m_{pl}^4$ and $\xi=1.0$, and the
initial conditions are $\phi_i=6.916\times10^{-4}m_{pl}$,
$\dot\phi_i=3.215\times10^{-4}m_{pl}^2$.}\label{linear2}
\end{figure}

In order to support our analytical calculation, we also try to find
some regions in parameter space which can give good numerical
results. In order to satisfy the constraints on the potential, we
take into account two forms of potentials: 1)
$V(\phi)=\frac{1}{2}m^2\phi^2-V_0$ and 2)
$V(\phi)=V_0(e^{\lambda\phi^2}-\frac{1}{2})$. We add some negative
zero-point energy to the potential to let it get some negative
region. For the coupling term, we choose
$F(\phi)=\kappa^2\xi\phi^2$, where $\xi$ is some coupling constant.
In Figs. \ref{linear1} and \ref{linear2} we show some numerical
calculations on Hubble parameter $H$ and scale factor $a$. We can
see that a bounce can happen naturally with $H$ crosses the zero
divide line. This indicates that by this way, the Big Bang
Singularity can be avoided.

\subsection{single scalar field with DBI form}
Beside canonical ones, there is another form of scalar field that is
very common used in cosmology: the Dirac-Born-Infeld (DBI) form.
Actions of this form effectively describes tachyon dynamics, which
can be obtained naturally in string theory \cite{Leigh:1989jq}.
Moreover, it can also drive the acceleration of the universe and can
act as inflaton \cite{Dvali:1998pa} and dark energy
\cite{Gibbons:2002md}. As inflationary models inspired by the string
theory, the DBI action non-minimally coupled to gravity has also
been investigated in the literature, cf. Ref. \cite{Easson:2009kk}.
Here we study the last case to see whether it can avoid the Big-Bang
Singularity.

The DBI lagrangian containing non-minimal coupling term is:
\be\label{dbi} P(X,\phi)=-V(\phi)\sqrt{1-2\alpha X}~,\ee where
$\alpha$ is some positive constant. Its energy density can be
calculated as $\rho=2XP_X-P=\frac{V(\phi)}{\sqrt{1-2\alpha X}}$. One
can also obtain its equation of state and sound speed square by
their definitions: \be w\equiv\frac{P}{\rho}=2\alpha
X-1~,~~~~~c_s^2\equiv\frac{P_X}{\rho_X}=\sqrt{1-2\alpha X}~,\ee
which shows that this kind of action is stable under classical
perturbations $(0<c_s^2<1)$ with Quintessence-like behavior
($-1<w<0$).


From (\ref{dbi}) we obtain the equation of motion for the scalar
field: \be -\frac{\alpha V}{(1-2\alpha
X)^{\frac{1}{2}}}\Box\phi-\frac{\alpha^{2}V}{(1-2\alpha
X)^{\frac{3}{2}}}\dot{X}\dot{\phi}-\frac{\alpha V_{\phi}}{(1-2\alpha
X)^{\frac{1}{2}}}\dot{\phi^{2}}-V_{\phi}\sqrt{1-2\alpha
X}-\frac{RF_\phi}{2\kappa^2}=0~,\ee where we get: \be
\ddot{\phi}=-\frac{[2\kappa^2(3H\alpha
V\dot{\phi}+V_{\phi})+RF_\phi(1-2\alpha X)^{\frac{1}{2}}](1-2\alpha
X)}{2\kappa^2\alpha V}\ee and the Friedmann equation: \bea
&&3H^{2}(1-F(\phi))=\kappa^{2}\frac{V(\phi)}{\sqrt{1-2\alpha X}}+3HF_{\phi}\dot{\phi}~,\\
&&\dot{H}=\frac{-\kappa^{2}[\kappa^{2}\frac{2\alpha^{2}V^{2}X}{(1-2\alpha
X)^{\frac{1}{2}}}+\alpha VHF_{\phi}\dot{\phi}+F_{\phi}(1-2\alpha
X)(3H\dot{\phi}\alpha V+V_{\phi})-2F_{\phi\phi}X\alpha
V]-6H^{2}F_{\phi}^{2}(1-2\alpha
X)^{\frac{3}{2}}}{2\kappa^{2}(1-F)\alpha V+3F_{\phi}^{2}(1-2\alpha
X)^{\frac{3}{2}}}~.\eea

From above we can see that the first condition at the bounce point
$\rho=0$ only requires $V=0$, which is looser than that of the
canonical scalar field case. Moreover, $F_\phi V_\phi<0$ is still
required by the second condition $\dot H>0$. Therefore, in order to
have a bounce, a region of negative value of $V(\phi)$ is also
inevitable.

\begin{figure}[htbp]
\includegraphics[scale=0.3]{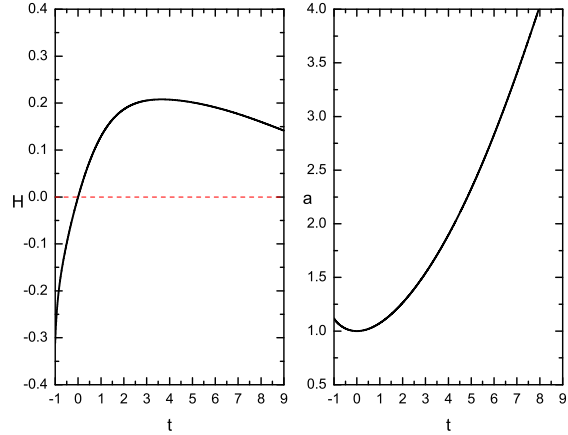}
\caption{The potential is chosen to be
$V(\phi)=V_0(e^{\lambda\phi^2}-\frac{1}{2})$ and the coupling term
is $RF(\phi)=\kappa^2\xi R\phi^2$. The parameter is chosen to be
$V_0=1.0m_{pl}^4$, $\alpha=1.0m_{pl}^{-4}$,
$\lambda=-1.0m_{pl}^{-2}$ and $\xi=1.0$, and the initial conditions
are $\phi_i=0.407m_{pl}$, $\dot\phi_i=0.776m_{pl}^2$.}\label{dbi3}
\end{figure}

\begin{figure}[htbp]
\includegraphics[scale=0.3]{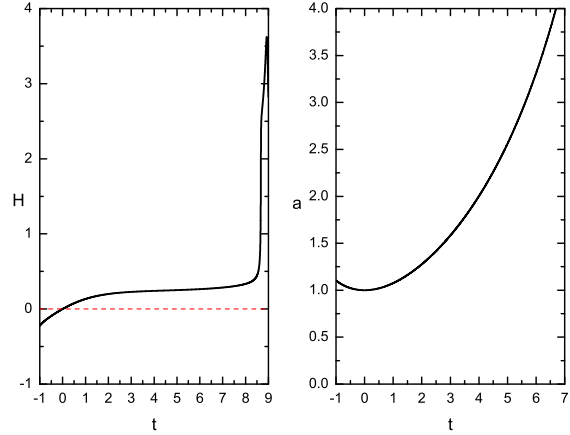}
\caption{The potential is chosen to be
$V(\phi)=V_0(e^{\lambda\phi^2}-\frac{1}{2})$ and the coupling term
is $RF(\phi)=\kappa^2\xi R\phi^2$. The parameter is chosen to be
$V_0=1.0m_{pl}^4$, $\alpha=1.0m_{pl}^{-4}$,
$\lambda=-1.0m_{pl}^{-2}$ and $\xi=1.0$, and the initial conditions
are $\phi_i=0.500m_{pl}$, $\dot\phi_i=0.548m_{pl}^2$.}\label{dbi2}
\end{figure}

We also make the numerical calculations, with the potential of the
form $V(\phi)=V_0(e^{\lambda\phi^2}-\frac{1}{2})$ and some proper
parameter choice. The bounce happens naturally from the view of the
plots. As a side remark, one may notice that due to different
initial conditions, the model may present various behaviors. From
Fig. (\ref{dbi3}) we can see that, the hubble parameter varies very
fast after the bounce, from increasing to decreasing, indicating
that the universe will enter into a moderate accelerating or
decelerating expansion phase soon. While from Fig. (\ref{dbi2}), one
may note that the hubble parameter experienced a period of slow
variation. This indicates that it is also possible to give rise to
an inflationary period after bounce, although we will not discuss in
detail as it goes beyond current topic.

\section{Category II: Gravity couples non-linearly to the field}
As a comparison, in this section, we will focus on a more general
case where, in the coupling term, Gravity possesses a non-linear
form. The most general form of basic equations comes from Eq.
(\ref{fe1}-\ref{eom}), However, for the sake of simplicity, we only
discuss about the canonical field (\ref{canonical}). Therefore, we
can derive the Friedmann equations and the equation of motion
explicitly: \bea \ddot
H&=&\frac{1}{18Hf_{RR}}[(Rf_R-f)/2-3Hf_{R\phi}\dot\phi+\kappa^2(\frac{{\dot\phi}^2}{2}+V(\phi))-3f_RH^2]-4H\dot H~,\\
\dddot H&=&\frac{1}{6f_{RR}}[H(f_{RR}\dot
R+f_{R\phi}\dot\phi)-2f_R\dot H-\kappa^2\phi^2-f_{RRR}{\dot
R}^2-2f_{RR\phi}\dot\phi\dot
R-f_{R\phi\phi}{\dot\phi}^2-f_{R\phi}\ddot\phi]-4H\ddot H-4{\dot
H}^2~,\eea where \be
\ddot\phi+3H\dot\phi+V_\phi-\frac{f_\phi}{2\kappa^2}=0~.\ee

Since the non-linear term of Gravity has been involved, the higher
order derivatives of $H$ appears in the equation, which makes the
equations difficult to solve analytically. But as the order of
derivative increases, the number of effective degrees of freedom
becomes more, and it will be easier to violate NEC and realize the
bouncing process.

\begin{figure}[htbp]
\includegraphics[scale=0.3]{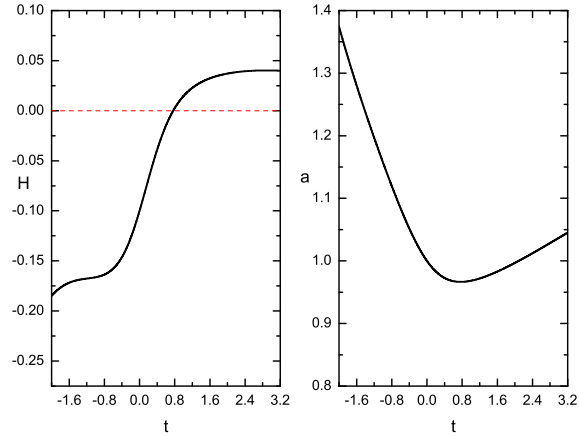}
\caption{The potential is chosen as $V(\phi)=\frac{1}{2}m^2\phi^2$
and the coupling term is chosen as
$f(R,\phi)=R-2\kappa^2\xi\ln{(\frac{R}{R_0})}\phi^4$. The parameter
is chosen to be $m=1.0m_{pl}$, $\xi=1.0$ and $R_0=1.0m_{pl}^{2}$,
and the initial conditions are $\phi_i=0.783m_{pl}$,
$\dot\phi_i=-0.205m_{pl}^2$.}\label{log3}
\end{figure}

\begin{figure}[htbp]
\includegraphics[scale=0.3]{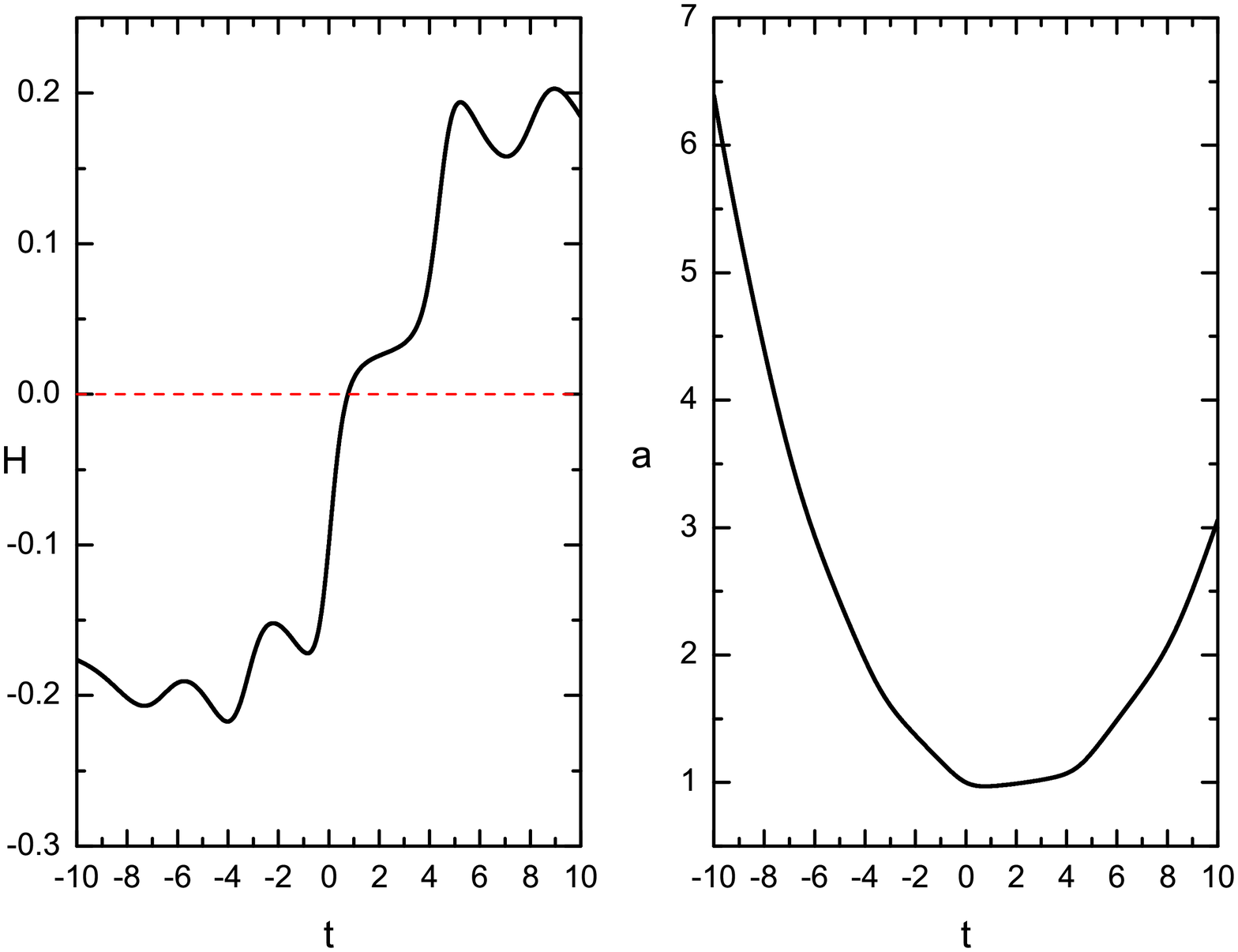}
\caption{The potential is chosen as $V(\phi)=\frac{1}{2}m^2\phi^2$
and the coupling term is chosen as
$f(R,\phi)=R-2\kappa^2\xi\ln{(\frac{R}{R_0})}\phi^4$. The parameter
is chosen to be $m=1.0m_{pl}$, $\xi=1.0$ and $R_0=1.0m_{pl}^{2}$,
and the initial conditions are $\phi_i=0.500m_{pl}$,
$\dot\phi_i=0.036m_{pl}^2$.}\label{log4}
\end{figure}

\begin{figure}[htbp]
\includegraphics[scale=0.3]{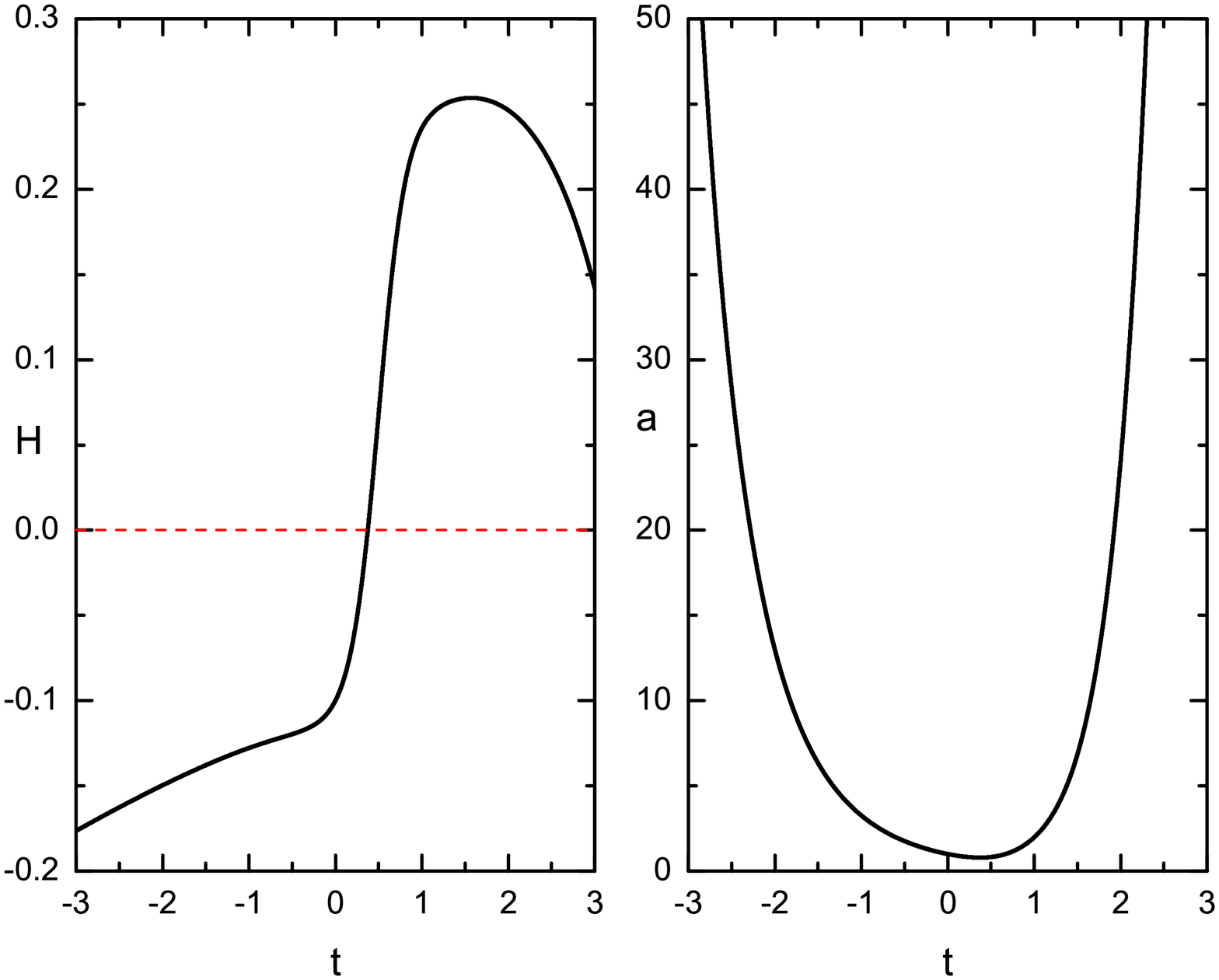}
\caption{The potential is chosen as $V(\phi)=\frac{1}{2}m^2\phi^2$
and the coupling term is chosen as $f(R,\phi)=R-2\kappa^2\xi
R^2e^{\lambda\phi}$. The parameter is chosen to be $m=0.1m_{pl}$,
$\xi=0.5$ and $\lambda=-5\times 10^{-3}m_{pl}^{-1}$, and the initial
conditions are $\phi_i=0.621m_{pl}$,
$\dot\phi_i=-0.039m_{pl}^2$.}\label{exp2}
\end{figure}

Figs. \ref{log3} and \ref{log4} show that a canonical field with
coupling term contains logarithm function of $R$. The logarithm
coupling may seem strange, however, one can find the similar form in
previous phenomenological studies, see e.g. \cite{Feng:2004mq}. We
can see from the plot that with proper choice of parameter, this
kind of coupling may also give rise to a bounce scenario. Fig.
\ref{exp2} is the plot of universe behavior for the coupling of
$R^2$ to some exponential potential of $\phi$. This kind of
potential looks like the dilaton potential in string theory
\cite{Gasperini:1992em}. With this kind of coupling, the universe
can also pass through the bouncing point smoothly.

\section{conclusions and discussions}
In this note we investigated the possibilities of a single scalar
field giving rise to a bouncing scenario in the very early universe,
with some non-minimal coupling to gravity. Along with some explicit
examples, we showed that it is quite possible. However, for the
common case that Gravity couples linearly to the field, the field
will suffer from some severe conditions, such as abandoning the
positive-definition of the potential. For other cases where Gravity
couples non-linearly, it will be more free to get a bounce since
more degrees of freedom have been evoked.

The realization of bounce with non-minimal coupling field is
important in the sense that, with the involvement of Gravity, one no
longer needs any ghost field to violate NEC, which might cause
problems. Recently there have been many works on construction of
bouncing cosmologies within the framework of non-minimal coupling
theories, which have some relation to our work while focus more on
their fundamental origins \cite{Setare:2008qr}. Furthermore, it is
expected that with proper choice of the coupling form, it is also
possible to obtain right amount of observational signatures to meet
the data, and to give rise to some new features and predictions for
future experiments, for example, a recent work shows that the matter
bounce scenario with non-minimal coupling will give rise to
scale-invariant spectrum and large particle productions
\cite{Qiu:2010ch}. All these fancy topics are left for the
forthcoming work.
\section*{Acknowledgments}

The author thanks Kazuharu Bamba, Prof. Je-An Gu, Prof. Anupam
Mazumdar, Prof. Yunsong Piao and Yi Wang for useful suggestions at
the beginning of the work. The research is supported in parts by the
National Science Council of R.O.C. under Grant No.
NSC96-2112-M-033-004-MY3 and No. NSC97-2811-033-003 and by the
National Center for Theoretical Science.

\end{document}